\documentclass{iopart}

\pdfoutput=1

\usepackage[english]{babel}
\usepackage[utf8]{inputenc}
\usepackage[colorlinks=true,linkcolor=blue, citecolor=blue, urlcolor=blue, bookmarks]{hyperref}
\usepackage{slashed}
\usepackage{amssymb}
\usepackage{tikz}

\def\sign{{\rm sign}}
\def\beq{\begin{equation}}
\def\eeq{\end{equation}}
\def\barray{\begin{eqnarray}}
\def\earray{\end{eqnarray}}

\def\R{{\mathbb R}}

\def\ket|#1>{| #1 \rangle}
\def\bra<#1|{\langle #1 |}
\def\<{\langle}
\def\>{\rangle}
\def\{{\lbrace}
\def\}{\rbrace}
\def\({\left(}
\def\){\right)}
\def\[{\left[}
\def\]{\right]}

\begin{document}

\title{More on the rainbow chain: entanglement, space-time geometry and thermal states}

\author{Javier Rodr\'{\i}guez-Laguna$^1$, J\'er\^ome Dubail$^2$, Giovanni Ram\'{\i}rez$^3$, Pasquale Calabrese$^4$,
Germ\'an Sierra$^5$}

\address{$^1$ Dto. F\'{\i}sica Fundamental, Universidad Nacional de  Educaci\'on a Distancia (UNED), Madrid, Spain.\\
$^2$ CNRS \&\ IJL-UMR, Universit\'e de Lorraine, F-54506 Vandoeuvre-les-Nancy, France.\\
$^3$ Instituto de Investigaci\'on, Escuela de Ciencias F\'{\i}sicas y Matem\'aticas, Universidad de San Carlos de Guatemala, Guatemala\\
$^4$ SISSA and INFN, Via Bonomea 265, 34136 Trieste, Italy \\
$^5$ Instituto de F\'{\i}sica Te\'orica (IFT), UAM-CSIC, Madrid, Spain
}

\begin{abstract}
The rainbow chain is an inhomogenous exactly solvable local spin model that, in its ground state, displays a half-chain 
entanglement entropy growing linearly with the system size. 
Although many exact results about the rainbow chain are known, the structure of 
the underlying quantum field theory has not yet been unraveled. 
Here we show that the universal scaling features of this model are captured 
by a massless Dirac fermion in a curved space-time with constant negative curvature $R=-h^2$ 
($h$ is the amplitude of the inhomogeneity).
This identification allows us to use recently developed techniques to study inhomogeneous conformal systems
and to analytically characterise the entanglement entropies of more general bipartitions. 
These results are carefully tested against exact numerical calculations.
Finally, we study the entanglement entropies of the rainbow chain in thermal states, and find that there is 
a non-trivial interplay between the rainbow effective temperature $T_R$ and the physical temperature $T$.
\end{abstract}

\maketitle

\section{Introduction}

In recent years there has been an intense activity to study extended quantum systems from the perspective of 
entanglement \cite{A08,E10,rev,rev-lafl}. 
This concept is certainly not new but it is moving to central stage in several areas of physics such as condensed matter 
physics, quantum optics, quantum field theory and quantum gravity. 
Furthermore the entanglement entropies recently became accessible to cold atomic experiments \cite{islam,kauf} 
and there are a number of theoretical proposals to access many other universal features of 
entanglement \cite{c-11,ad-12,dpsz-12,pzszh-16,a-16,uzpy-16}.
Consequently, it is of fundamental importance to characterise the entanglement properties of
many-body systems, especially in those situations that can be realised experimentally. 
More and more attention has been recently drawn to inhomogeneous many-body systems in one spatial dimension 
for a twofold reason: on the one hand, in nature inhomogeneities are more the rule rather than the exception, 
on the other hand, in some of these models the entanglement properties exhibit interesting departures from the laws
satisfied by uniform models. 
For example, spin chains with quenched disorder in the couplings are not conformal invariant but show, in average, logarithmic
entanglement entropies \cite{RM04,RM04b,L05,FCM11,RRS14,rac-16}, 
similarly to the uniform conformal case \cite{H94,V03,CC04,cc-09} but with a different prefactor.  
Lattice models with couplings that increase exponentially with the position show analogies with Wilson's RG approach
to the Kondo problem \cite{ON10}. Hyperbolic couplings have been
introduced to reduce the effect of the boundary in estimating the
energy of the excited states that become weakly bounded around the
centre of the system \cite{UN09,UNKN10}. 
Finally, and more importantly for this manuscript, spin chains with couplings that decrease
exponentially moving away from the centre of the system give rise to an entanglement scaling with volume of the subsystem 
even in the ground state \cite{VRL10,RRS14b,RRS15}, resulting in a very anomalous behaviour compared to uniform models.

The similarity between models with smoothly varying couplings and
quantum field theories in curved space-times has been put forward in
Refs. \cite{BCLL11,RTLC16}, with potential cold atom realisations in
sight. In addition to those proposals for tabletop experiments that
mimic effects from high-energy physics, there is another good reason
for exploring the relevance of quantum field theory in curved space for
the physics of ultracold quantum gases.  Most experimental setups
involve trapping potentials, often harmonic ones, that result in
inhomogeneous density profiles in the models one wants to simulate; in
fact, the presence of non-uniform potential wells, or background
electromagnetic fields, is the rule rather than the exception. It was
pointed out very recently \cite{DSVC16}, in the study of
non-interacting fermion gases in 1+1 dimensions, that the
inhomogeneity generated by the trapping potential are captured by
quantum field theory in curved space-time.

It is thus of great interest to find a general framework where we can
study the entanglement properties of those inhomogeneous systems. The
goal of this paper is to apply the techniques of Ref. \cite{DSVC16} to
the {\em rainbow} model \cite{VRL10,RRS14b,RRS15}.
In this way, we will understand its entanglement properties and its geometrical significance. 
The paper is organised as follows. In section
\ref{sec:review} we review the main properties of the rainbow chain.
In section \ref{sec:dirac_cst} we carry out the continuum limit of the rainbow model in
its Lagrangian formulation, studying in full  details the resulting space-time geometry.
In section \ref{sec:entropies} we provide the general CFT expressions of the R\'enyi entanglement entropies for an arbitrary block 
of spins, both in the case of a block starting from one end of the chain  and of a block with endpoints in the bulk of the system. 
The corresponding analytical expressions are carefully verified numerically for several system sizes and inhomogeneity amplitudes, 
providing in some cases also conjectures for the leading correction to scaling.  
In section \ref{sec:temp} we numerically study the entanglement entropy at non-zero temperature finding an interesting interplay with
the effective rainbow temperature.  
Finally, in section \ref{ref:conclusions} we present our conclusions and some possible prospects.


\section{Review of the rainbow chain}
\label{sec:review}

\begin{figure}[t]
  \centering
  \includegraphics[width=\textwidth]{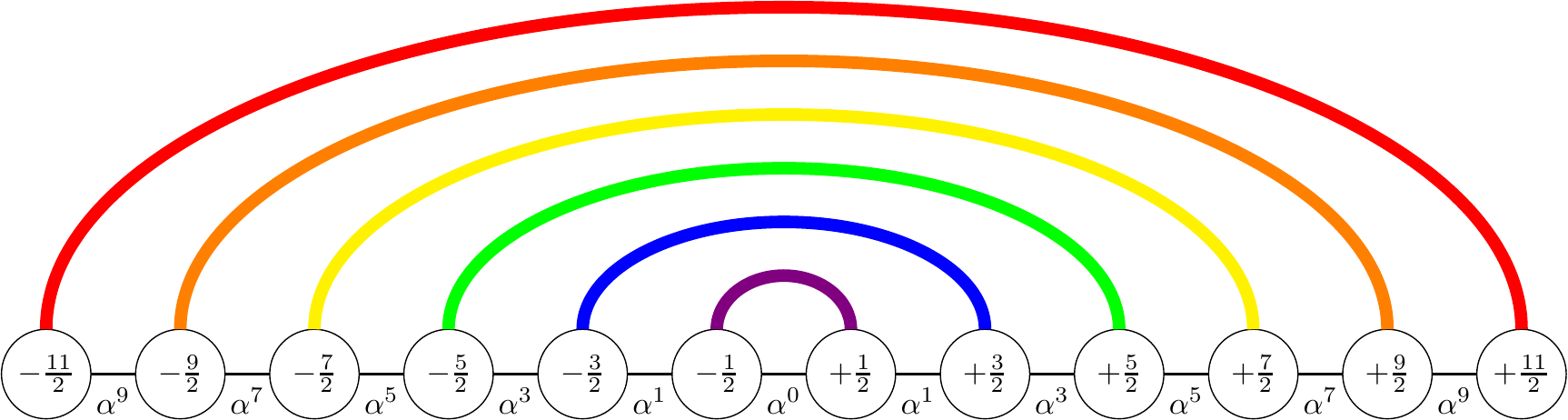}
  \caption{Rainbow state showing the $(-k,+k)$ bonds above the central
    link. Given that each bond contributes as $\ln2$ to the entanglement entropy, the one between the left and the
    right halves of the chain is $L \ln 2$.}
  \label{fig:RB_scheme}
\end{figure}

To define the rainbow model, we consider a chain with $2L$ sites
labelled by the half-odd integers, $m=\pm\frac{1}{2}, \pm \frac{3}{2},
\cdots, \pm\(L-\frac{1}{2}\)$. Let $c_m$ and $c^\dagger_m$ denote the
annihilation and creation operators of a spinless fermion at the site
$m$. The Hamiltonian is  \cite{VRL10,RRS14b,RRS15}
\beq
  H =  -\frac{J}{2} c_{\frac{1}{2}}^\dagger c_{-\frac{1}{2}} - 
  \frac{J}{2}  \sum_{m=\frac{1}{2}}^{L -\frac{3}{2}} e^{-hm}  
\[c^\dagger_m c_{m+1} +c^\dagger_{-m} c_{-(m+1)}\] +\mathrm{h.c.}\,,
  \label{1}
\eeq
where $J>0$ sets the scale of the hopping parameters, and $h \geq 0$
characterises the inhomogeneity of the hopping amplitudes. After a
Jordan-Wigner transformation Eq. (\ref{1}) becomes the Hamiltonian of
an inhomogeneous spin-1/2 $XX$ chain \cite{VRL10}. 
The case $h=0$ corresponds to the standard uniform Hamiltonian of a spinless free fermion 
(tight binding model) with open boundary conditions. 
Its low energy properties are captured by a CFT with central charge $c=1$: the massless Dirac fermion
theory, or equivalently (upon bosonization) a Luttinger liquid with Luttinger parameter $K=1$. 
The ground state of the Hamiltonian  (\ref{1}) has been studied in the strong and
in the weak inhomogeneity limits \cite{VRL10,RRS14b,RRS15}. In the
former case, i.e. $h \gg 1$, the application of the Dasgupta-Ma real
space renormalisation group \cite{Ma.PRB.80} yields a ground state
formed by spin singlets between the sites $k$ and $-k$, for $k= \frac{1}{2}, \dots, L-\frac{1}{2}$ (in the $XX$ version of the
Hamiltonian).  Fig. \ref{fig:RB_scheme} shows a picture of the ground state that
resembles a rainbow (the colours are associated to the energies
required to break the bonds, which are proportional to $e^{-2mh}$, and
so red shifting with $|m|$).  

In this paper we will be mainly interested in the characterisation of the block entanglement entropies in the ground state of the 
rainbow chain (\ref{1}). 
Given a block $A$ formed by $\ell$ contiguous spins, the R\'enyi entanglement entropies  are defined as 
\begin{equation}
S_n = \frac{1}{1-n}\ln {\rm Tr} ( \rho_A^n),
\end{equation}
where $\rho_A\equiv  {\rm Tr}_{\bar A} \rho $ is the reduced density matrix of  $A$ and $n$ is the order of the R\'enyi 
entropy which can be an arbitrary real number. 
For $n\to1$, $S_n$ reduces to the von Neumann entropy of the subsystem, which is usually referred to as entanglement entropy, and 
that will be the main quantity of interest also for us.
In the following we will simply refer to $S_1$ as $S$, i.e. every time that the R\'enyi index is missing, we implicitly 
assume to consider the von Neumann entropy.
It is however important to characterise the R\'enyi entropies for general $n$, indeed their
knowledge for arbitrary $n$ gives access to the full spectrum of the reduced density matrix \cite{cl-08}.
Given that the rainbow chain is a free fermionic model, one can use standard techniques \cite{ffo} to evaluate the 
entanglement entropies for finite (but also very large) $L$, as already done in  \cite{VRL10,RRS14b,RRS15}.

A fundamental property of the entanglement entropy is that, in the ground state of gapped models, it satisfies the so-called area 
law \cite{E10,bombelli,area-law}, i.e. when increasing the subsystem size it scales like the area of $A$, in contrast with the volume law of standard 
thermodynamic entropy at finite temperature. 
In one spatial dimension, the area law means that the entanglement entropies saturate when increasing the block size; 
this was proven rigorously by Hastings \cite{H07}.
%


For gapless systems,  there are many known fundamental examples where the area law breaks down, including logarithmic violations in conformal field theories \cite{H94,V03,CC04,cc-09},  systems with quenched disorder \cite{RM04,RM04b,L05,FCM11,RRS14,rac-16}, 
Fermi gases in arbitrary dimension \cite{Wolf-06,GK-06,cev-12} and many more. 
Conversely, an unexpected result was the one by Vitigliano {\it et al.} \cite{VRL10} 
showing that the rainbow state (i.e. the ground state of the rainbow model) breaks maximally the area law in the limit $h \gg1$. 
Indeed, since any singlet shared between one block and its complement contribute to the 
entanglement entropy as $\ln2$, it is clear from Fig. \ref{fig:RB_scheme} that the half-chain entanglement
entropy (i.e. between the left and right blocks) of this state is $S=L\ln 2$, which is the highest value it can take for a subsystem containing $L$ qubits. 
This valence bond state becomes the exact ground state of the
Hamiltonian  (\ref{1}) in the limit $h \rightarrow \infty$, but several
of its properties persist for all values of $h$, in particular the
linear dependence on $L$ of the half-block entropy with a coefficient 
that depends on $h$ and it is always smaller than $\ln 2$.

\begin{figure}[t]
\centering
\includegraphics[width=.8\textwidth]{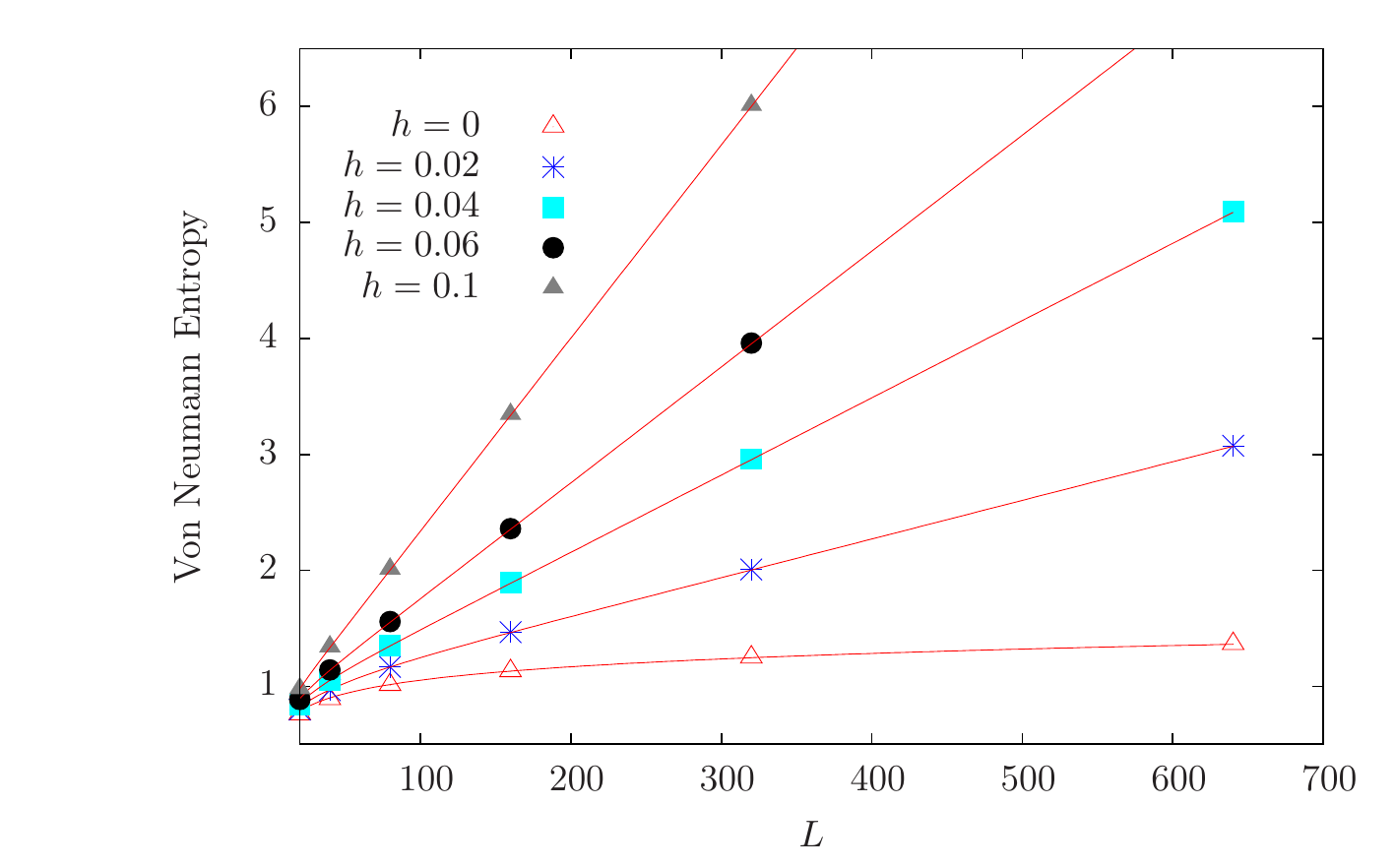}
\caption{
Half-chain entanglement entropy for the rainbow model. 
The figure reports the comparison between exact numerical data (symbols) and the theoretical prediction (\ref{10}) (lines)
for different values of $L$ and $h$. The agreement is perfect in the considered regime of small $h$.}
\label{fig:S_prediction}
\end{figure}

In the weak inhomogeneity limit $h \ll 1$, it has been shown that the low energy physics of
 (\ref{1}) is described by the Hamiltonian of two chiral fermions
$\psi_L$ and $\psi_R$ of the form \cite{RRS15}
\beq
\fl H  \simeq    i  J a \int_{-a  L}^{a L} dx \,  e^{-\frac{h |x|}{a}}
\[ \psi^\dagger_R\partial_x\psi_R - \psi^\dagger_L\partial_x\psi_L
- \frac{h}{2 a} \sign{(x)} (\psi^\dagger_R\psi_R - \psi^\dagger_L\psi_L) \] ,
\label{2} 
\eeq
where $a$ is the lattice spacing and $x=ma$ is the position. The
fields $\psi_{L,R}(x)$ are the slow varying modes of the fermion
operator $c_m$ expanded around the Fermi momenta $\pm k_F$ 
(at half filling $k_F=\pi/(2a)$)
\beq
\frac{c_m}{\sqrt{a}} \simeq e^{i k_F x} \psi_L(x) +e^{-i k_F x} \psi_R(x).
\label{3}
\eeq 
The continuum limit is defined by the equations $a\to 0$, $h\to 0$ and
$L\to\infty$, with $h/a$ and $aL$ kept constant. In the rest of the
paper, we rename $h/a \rightarrow h$ and $aL \rightarrow L$ such that
$h$ and $L$ acquire the dimensions of length$^{-1}$ and length
respectively. In Eq.  (\ref{2}), the fields $\psi_{L,R}$ are decoupled
in the bulk, however they are coupled by the boundary conditions
\cite{RRS15}
\beq
\psi_R (\pm L) = \mp i \,  \psi_L (\pm  L).
\label{4}
\eeq
In reference \cite{RRS15} it was noticed that Eq.  (\ref{2}) can be
transformed into the Hamiltonian of a uniform model by the change of variable
\beq
\tilde x = \sign{(x)} \frac{e^{ h |x|} -1}{h},
\label{5}
\eeq
that maps the interval $x \in [-L, L]$ into the interval $\tilde{x}
\in [-\tilde{L},\tilde{L}]$ where
\beq
\tilde{L} = \frac{e^{h L} -1}{h},
\label{6}
\eeq
plays the role of an effective half-length. The fermion fields in the
variable $\tilde x$ transform as
\beq
\tilde{\psi}_{R,L} (\tilde{x})  =  \( \frac{d\tilde{x} }{d x} \)^{-1/2}
\psi_{R,L}(x) = e^{-h |x|/2} \psi_{R,L}(x),
\label{7}
\eeq
that plugged into Eq.  (\ref{2}) gives
\beq
  H \simeq i J \int_{- \tilde L}^{\tilde L} d \tilde{x} \, \[
  \tilde{\psi}^\dagger_R \partial_{\tilde x} \tilde{\psi}_R
  -\tilde{\psi}^\dagger_L \partial_{\tilde{x}} \tilde{\psi}_L \].
  \label{8}
\eeq 
The change of variables  (\ref{7}) also changes the integration measure
of the path integral, so that the transformed fermion is not living anymore in flat space-time, unlike  Eq.  (\ref{8}) 
could erroneously suggest.
Nevertheless, this equation was used heuristically in \cite{RRS15} to propose an expression for the half-chain entanglement
of the rainbow chain. The trick was to take the entanglement entropy
of a homogeneous open chain in CFT \cite{CC04}
\beq
S_{\rm CFT}(L) = {c\over 6} \ln(L) + c',
\label{9}
\eeq
with $c=1$, and replace  $L$ by $\tilde{L}$, to obtain 
\beq 
S(L) = {c\over 6} \ln\( {e^{hL}-1\over h} \) + c' \, . 
\label{10}
\eeq
A comparison of this prediction with numerical data is displayed in Fig. \ref{fig:S_prediction}, for small values of $h$, 
and the agreement is perfect. 

In the limit $hL \ll 1$, Eq.  (\ref{10}), reproduces the usual CFT
behaviour  (\ref{9}), but for $hL\gg 1$ the entropy behaves extensively
\beq
S (L) \approx {c\over 6} hL,
\label{11}
\eeq
in close analogy with the thermal behaviour of the entanglement entropy
of an open system in CFT \cite{CC04}
\beq
  S_{\rm CFT}(L, T)\approx \frac{c}{6} \ln\( \frac{ \sinh \(2 \pi L T   \)}{\pi T} \) \to \frac{c}{3} \pi T L  ,
  \label{12}
\eeq
in the limit where $TL\gg 1$. The comparison between
Eqs.  (\ref{11}) and  (\ref{12}) suggests (in the limit $hL\gg1$) a thermal interpretation of
the rainbow state with an effective temperature proportional to $h$
\beq
T_R =\frac{h}{2\pi}  
\label{13}
\eeq
for $T_RL\gg 1$.
%




\section{The Rainbow model as a Dirac fermion in curved space-time} 
\label{sec:dirac_cst}

In this section we show that the rainbow Hamiltonian can be obtained from the action of a massless Dirac fermion in a curved
space-time. We provide a derivation of this result using the covariant formalism of relativistic field theory with 
Minkowski signature. The Lagrangian associated to the Hamiltonian
 (\ref{2}) is
\barray
{\cal L} &= & \psi^\dagger_- \partial_0 \,\psi_- +  
 \psi^\dagger_+ \partial_0 \,\psi_+ + 
 e^{ -h|x^1|} \left( \psi^\dagger_-  \partial_1 \,\psi_- - 
 \psi^\dagger_+ \partial_1 \,\psi_+ \right)  \nonumber\\
  &&  +      
 \frac{h}{2} {\rm sign}(x^1) \; e^{-h|x^1|}   
 ( \psi^\dagger_+ \psi_+ -  \psi^\dagger_-  \psi_- ) ,
\label{d1} 
\earray 
where $\psi_{-} = \psi_{L}$, $\psi_+= \psi_R$, $(t,x)= (x^0, x^1)$ and
$\partial_\mu = \partial/\partial x^\mu$.  For $h=0$, Eq. (\ref{d1})
becomes the Dirac action of a massless fermion in 1+1 dimensions, i.e. 
\barray 
{\cal L} &= & \,\bar{\psi}\,{\slashed\partial}\,\psi =
\psi^\dagger_- ( \partial_0 + \partial_1) \psi_- + 
\psi^\dagger_+ ( \partial_0 - \partial_1) \psi_+   ,
 \label{d2}  
\earray
where ${\slashed\partial} = \gamma^\mu \partial_\mu$, $\psi^t =
(\psi_-,\psi_+)$, $\bar{\psi} = \psi^\dagger \gamma^0$ and
$\gamma^0=\sigma^x$, $\gamma^1=-i\sigma^y$. The flat space-time
metric, denoted by $\eta^{a b} \; (a,b=0,1)$, has the signature
$(-,+)$. Let us express Eq.  (\ref{d1}) as the Dirac action in curved
space-time. We need the space-time metric $g_{\mu \nu}$, the zweibein
$e_\mu^a$, its inverse $E^\mu_a$, the spin connection
$\omega_\mu^{ab}$ and the Christoffel symbols
$\Gamma_{\lambda\mu}^\nu$ that are related as
\barray 
g_{\mu \nu} &= & \eta_{ab} e_\mu^a \, e_\nu^b, 
\quad 
E_a^\mu = g^{\mu\nu}\eta_{ab} e^b_\nu, \label{d3} \\
\omega_\mu^{a b} &= &  
e^a_\nu\,\partial_\mu E^{b\nu} +e^a_\nu E^{b \lambda} \Gamma_{\lambda\mu}^\nu\,. 
\nonumber 
\earray 
The covariant derivative on the two component spinor $\psi$ is given by
\beq
D_\mu \psi = \(\partial_\mu-\frac{1}{4} \omega_\mu^{ab} \gamma_{ab}\) \psi, 
\label{d4}
\eeq 
where $\gamma_{ab} = \frac{1}{2} [\gamma_a,\gamma_b] =
-\epsilon_{ab}\, \sigma^z$ ($\epsilon_{ab}$ is the Levi-Civita tensor,
$\epsilon_{01}=-\epsilon_{10}=1$). The Dirac Lagrangian of a massless
fermion in curved space-time reads
\barray
{\cal L}   & = &   \, e \, \bar{\psi}  {\slashed D}   \psi \; , 
\label{d5}
\earray 
where ${\slashed D} = E^\mu_a \gamma^a D_\mu$ and $e = \det
e_\mu^a$. Assuming that $e_\mu^a$ has only diagonal entries, and using
Eqs.  (\ref{d3}) one finds
\barray 
e\,\bar{\psi}{\slashed D} \psi  &= & 
e \psi^\dagger \[ E^0_0 \; (\partial_0+\frac{1}{2}\omega_0^{01}\,\sigma^z) \
+   E^1_1 \;(\sigma^z\partial_1+\frac{1}{2} \omega_1^{01}) \] \psi ,
\label{d6} 
\earray 
that compared with Eq. (\ref{d1}) yields
\barray 
e\, E^0_0 &= & 1, \quad e\,E^1_1 =e^{-h|x|}, \label{d7} \\
e\, E_0^0 \,\omega_0^{01} &= & -h\,{\rm sign}(x) \, e^{-h|x|}, 
\quad \omega_1^{01} =0 \,. 
\nonumber 
\earray 
The solution of these equations is  
\barray 
e_0^0 &= &   e^{-h|x|}, \quad e_1^1 = 1, \label{d8} \\ 
\omega_0^{01} &= & -h\,{\rm sign}(x)\, e^{-h|x|}, \quad 
\omega_1^{01} =0 \, , \nonumber 
\earray 
that gives rise to the space-time metric
\beq
g_{00} = -e^{-2h|x|}, 
\qquad 
g_{11} = 1 \,. 
\label{d9}
\eeq
The non vanishing components of the Christoffel symbols and the Ricci
tensor are
\barray 
\Gamma_{01}^0  &= & -h \,{\rm sign}(x), \quad 
\Gamma_{00}^1  = -h\, {\rm sign}(x)\, e^{-2h|x|}\, , \label{d10} \\
R_{00} &= & -2h\delta(x), \quad R_{11}=2h\delta(x)-h^2  \, , \nonumber   
\earray 
that yield the  scalar curvature 
\beq
R(x) = g^{\mu\nu} R_{\mu\nu}(x) = 4 h \delta(x) - h^2  \, . 
\label{d12}
\eeq
that is constant and negative everywhere except at the origin where it
is singular.

The Euclidean version of the metric  (\ref{d9}) is 
\beq
ds^2 = e^{2h|x|} dt^2 + dx^2 = e^{2\sigma(x)}dz \, d\bar{z},
\label{d91}
\eeq
that, using Eqs.  (\ref{5}) and  (\ref{17}), coincides with
Eq.  (\ref{16}).  An alternative way to write the metric is
\beq
ds^2 = \frac{ d\tilde{x}^2 + dt^2}{ ( 1 +  |h \tilde{x}|)^2} \, ,
\label{22}
\eeq
or, in terms of the variable
 \beq
y =  \tilde{x} + \frac{{\rm sign}\, \tilde{x}}{h} 
\longrightarrow   
ds^2 =  \frac{dy^2 + dt^2}{h^2 y^2},  \, \label{23}
\eeq
that is the Poincar\'e metric in the upper-half plane in the variable
is $w=y+it$. This result is in agreement with Eq.  (\ref{d12}) for the
value of the curvature. In our case, the domain where the metric is
defined consists in the union of two strips along the $t$ axis, i.e. 
\barray 
{\cal P}&= & {\cal P}_+ \cup {\cal P}_-, \quad 
{\cal P}_\pm = \R \times {\cal I}_\pm,  \label{d15} \\ 
{\cal I}_{\pm} &= & \pm(y_0,y_L), \qquad    
y_0=\frac{1}{h}, \quad 
y_L=\frac{e^{hL}}{h}  \nonumber   \, . 
\label{d15b}
\earray


\section{Entanglement entropies}
\label{sec:entropies}

The goal of this section is to show how to obtain the entanglement entropies of the rainbow chain 
with the methods of Ref. \cite{DSVC16}.
This will allow us to derive Eq. (\ref{10}) and to generalise it to many other different bipartitions. 

In the previous section we have shown that, in Lagrangian formalism (in the worldsheet
$(x,t) \in [-L,L] \times \mathbb{R}$, with {\em imaginary} time $t$),
the 2D euclidean action corresponding to the Hamiltonian  (\ref{2}) is 
\beq
{\cal S} = \frac{1}{2\pi} \int dz d \bar{z}  \; e^{ \sigma} \[ \psi^\dagger_R 
\stackrel{\leftrightarrow}{\partial_{\bar z}} \psi_R + 
\psi^\dagger_L \stackrel{\leftrightarrow}{\partial_z} \psi_L \] ,
\label{15}
\eeq
in the metric
\beq
ds^2  = e^{2 \sigma} dz \, d \bar{z} .
\label{16}
\eeq
The complex coordinate $z$, as well as the Weyl factor $e^{\sigma}$ follow from the previous
section, or can be read off directly from Eq.  (\ref{2}),
\begin{eqnarray}
  \label{17} &&
  z \,=\, \tilde{x} + i \, t \,
  = {\rm sign} (x) \frac{e^{h |x|}-1}{h} + i \,t   \, , \\
  && e^{\sigma} \, =\, e^{-h |x|}.
\nonumber 
\end{eqnarray}
The complex coordinate lives on the infinite strip, $z \in
[-\tilde{L},\tilde{L}] + i \mathbb{R}$. This is a particular example
of the general class of models considered in Ref. \cite{DSVC16}, so
one can follow the arguments used in there to calculate the entanglement entropy. 
We will make repeated use of the standard result that the $n^{\rm th}$
R\'enyi entropy can be expressed in terms of correlation functions of
twist operators in an $n$-times replicated worldsheet \cite{CC04}. 
The twist field has the well known dimension \cite{cc-09}
\beq
\label{eq:dimension}
\Delta_n= \frac{c}{12}\Big(n-\frac1n\Big),
\eeq
where in our case $c=1$.

\subsection{Block $A$ at the chain edge} 

We first consider the case of a block $A$ starting from the edge of the chain, i.e. we focus on the bipartition
\beq
A= \[-L,x\],  \quad B = \[x,L\] .
\label{18}
\eeq
The R\'enyi entropy in this situation is related to the expectation value of a single twist operator ${\cal T}_n$ at position 
$(x,0)\in [-L,L]\times\R$ as
\beq
S_{n, {\rm CFT}}(x)= \frac1{1-n}\ln\, \< {\cal T}_n(x,0) \>_{\rm  strip \; (curved)} \,,
\eeq
where $\< . \>_{\rm  strip \; (curved)}$ stands for the expectation value in the CFT with the metric (\ref{16}). Since
the twist field is a primary operator with scaling dimension (\ref{eq:dimension}), it transforms as
$\mathcal{T}_n \rightarrow e^{-\sigma \Delta_n} \mathcal{T}_n$ under the Weyl transformation
of the metric $e^{2 \sigma}  d z d\overline{z} \rightarrow dz d\overline{z}$. Thus, the expectation
value of $\mathcal{T}_n$ in curved space is related to the one in flat space through
\begin{equation}
	\< {\cal T}_n(x,0) \>_{\rm  strip \; (curved)} \, =\, e^{-\sigma \Delta_n} \< {\cal T}_n(\tilde{x},0) \>_{\rm  strip \; (flat)} .
\end{equation}
The latter expectation value is entirely fixed by conformal invariance (the
boundary condition for the fermions does not require insertions of boundary condition changing operators),
and is easily calculated by conformally mapping the strip $z \in [-\tilde{L},\tilde{L}] + i \mathbb{R}$ onto the upper half-plane 
$\mathbb{H}$, $z \mapsto e^{i \frac{\pi}{2 \tilde{L}}( z+\tilde{L})} \in \mathbb{H}$.
%
The result of this straightforward calculation is  
\barray 
S_{n, {\rm CFT}}(x) & = & \frac{n+1}{12 n}  \ln Y(x), \label{21} 
\earray 
where $Y(x)$ is
\beq 
\fl Y(x)  =   e^{\sigma} \, \frac{8 \tilde{L}}{\pi}
\sin \( \frac{\pi (\tilde{x} + \tilde{L})}{2\tilde{L}} \)  
=    8  e^{-h|x|} \frac{e^{hL}-1}{\pi h} 
\cos\( \frac{\pi}{2} \frac{e^{h|x|}-1}{e^{hL}-1} \)  \, .
\label{21a} 
\eeq
The factor $8$ appearing in this expression is due to our choice of normalisation; it agrees, in the limit $h \to 0$, with
Eq. (70) of Ref. \cite{FC11} that gives the R\'enyi entropies of the uniform XX chain with open BC's. 
The term $\frac{2 \tilde{L}}{\pi}
\sin(\frac{\pi (\tilde{x} + \tilde{L})} {2\tilde{L}}\, )$ is the
standard {\em chord distance} entering in all CFT one-point function on the infinite strip. 

%
The half-chain entropy conjectured in Eq.  (\ref{10}) can be obtained
from Eq.  (\ref{21}) setting $n=1$ and $x=0$.

\begin{figure}[t]
\centering
\includegraphics[width=0.48\textwidth]{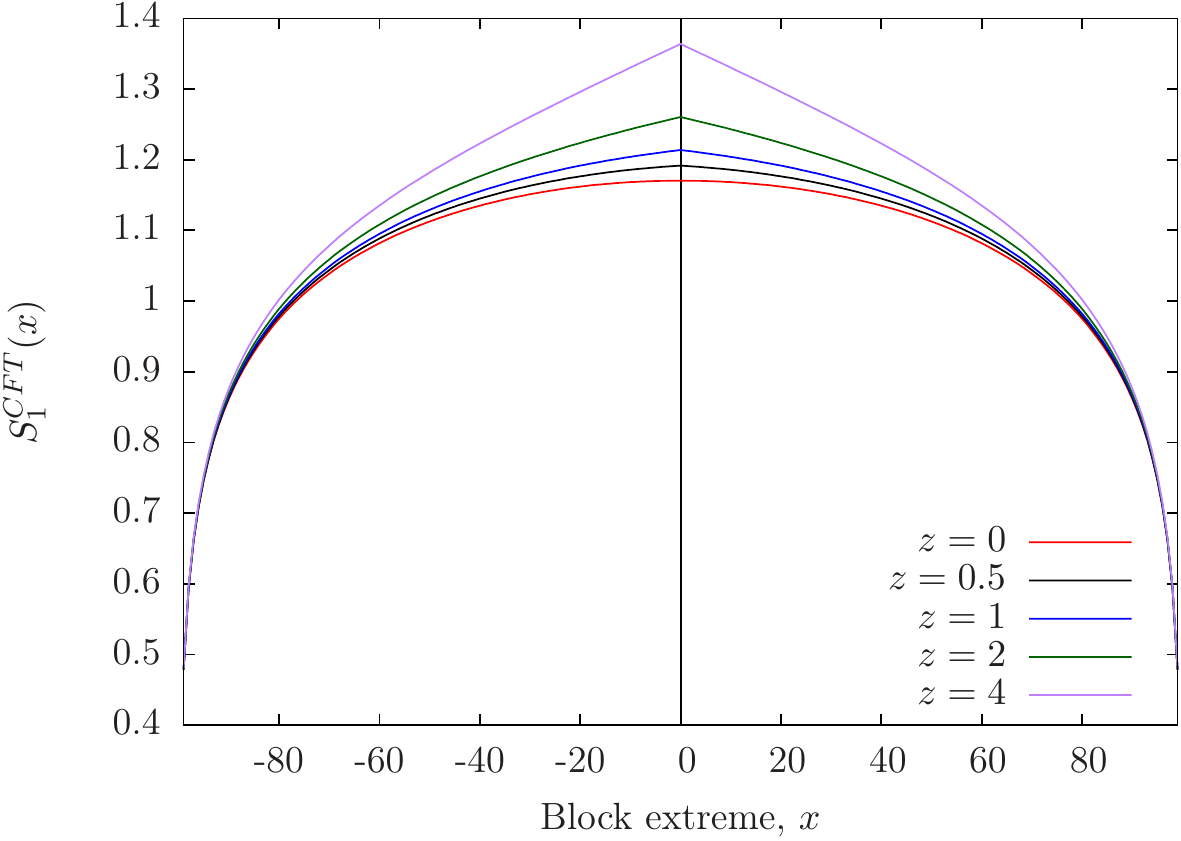}
\hspace{0.2cm}
\includegraphics[width=0.48\textwidth]{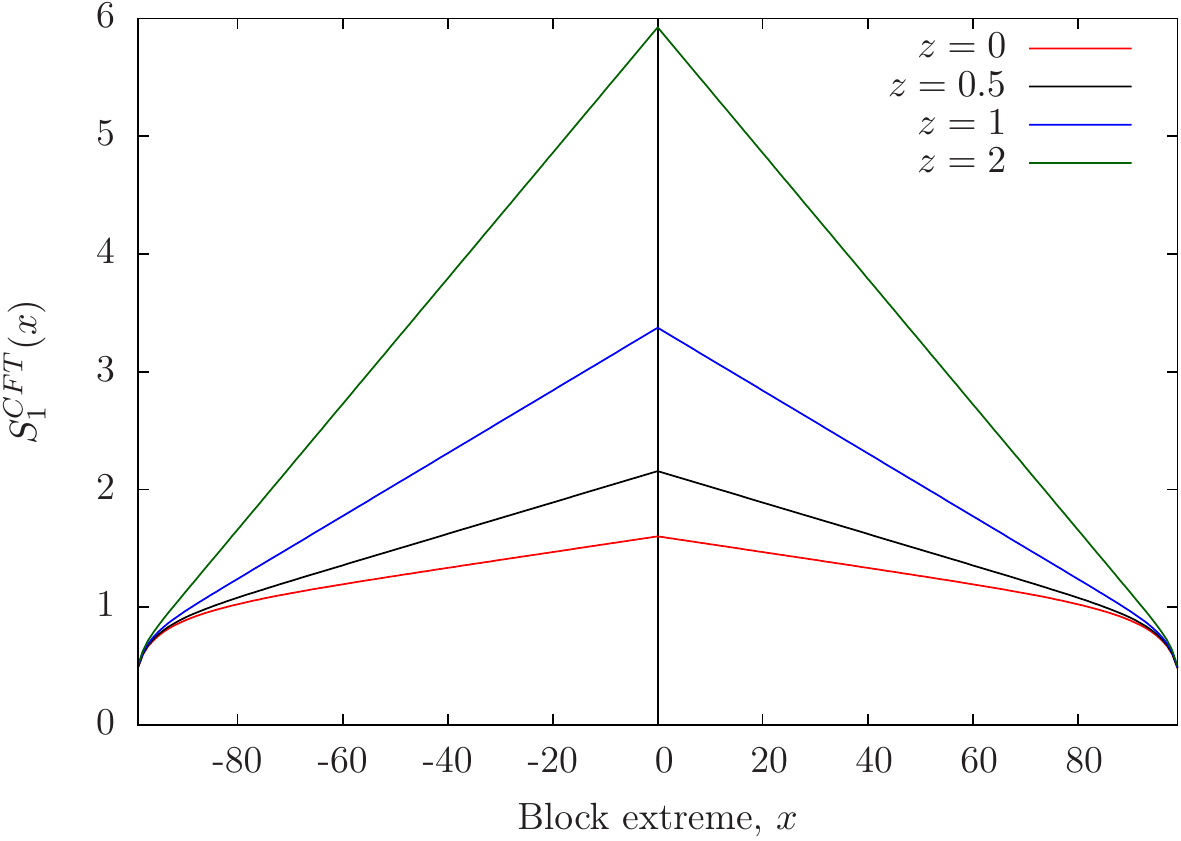}
\caption{CFT prediction for the entanglement entropy $S_1$ of the rainbow chain as given by  (\ref{21}) for $L=200$.
A part from the leading $\ln L$ behaviour, this is a function only of $x/L$ (horizontal axis) and $z=hL$. 
The latter is the varying parameter in the various curves which correspond to $z=0,0.5,1,2,4$ (left panel) 
and $z=8,16,32,64$ (right panel). 
The two panels shows neatly the crossover from the homogeneous CFT behaviour at $z=0$ to the thermal like 
at large $z$.
}
\label{fig:Sblock}
\end{figure}

Eq.  (\ref{21}) shows that, apart from the leading $\ln L$ behaviour, the R\'enyi entropies are just functions of $x/L$ and $z=hL$.
In Fig. \ref{fig:Sblock} we report this universal function for the von Neumann entropy (for other R\'enyi only the scale changes) 
as function of $x/L$ for several values of $z$. 
This figure shows the crossover from the homogeneous CFT behaviour at $z=0$ to the thermal like at large $z$.
Notice in particular that for $x$ close to the edges, one has always a logarithmic behaviour, but the region where 
this applies shrinks more and more as $z$ increases.

\subsubsection{Numerical tests.}

\begin{figure}[t]
\centering
\includegraphics[width=0.48\textwidth]{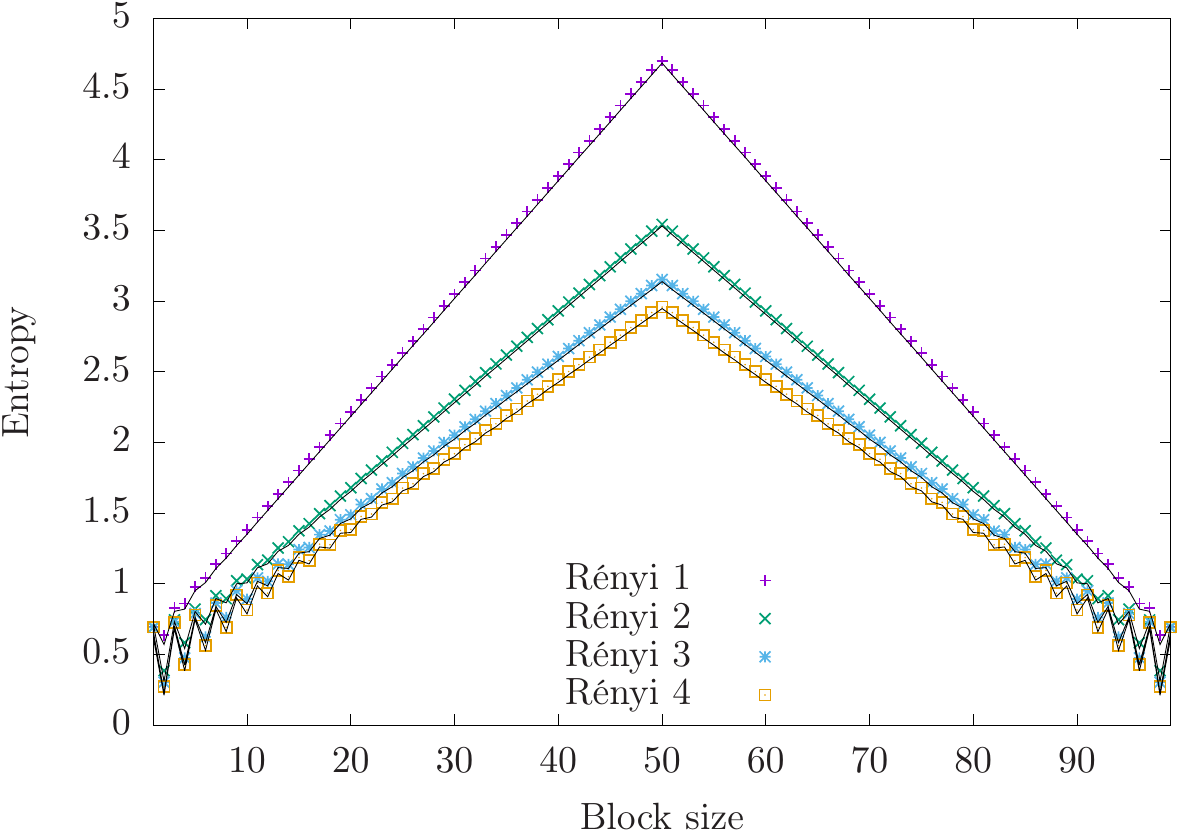}
\hspace{0.2cm}
\includegraphics[width=0.48\textwidth]{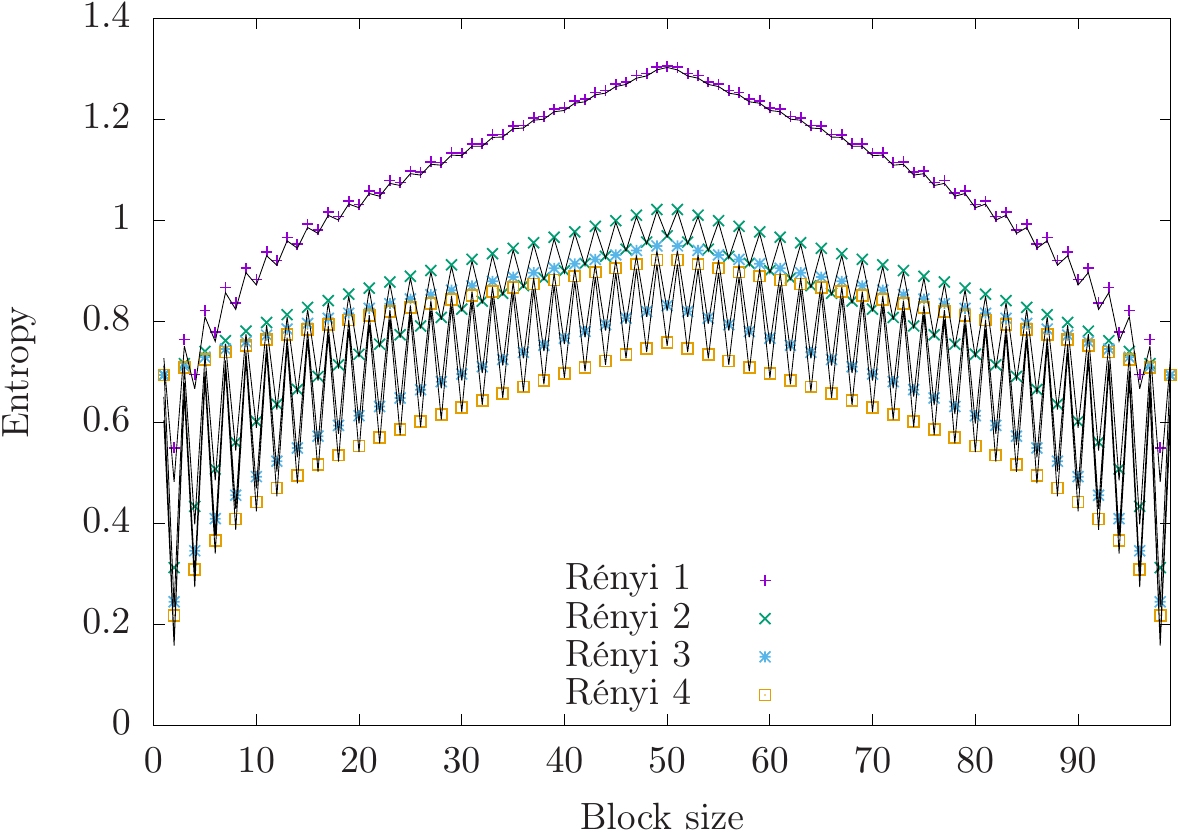}
\caption{R\'enyi entropies $S_n(x)$ for a system with $2L=100$ sites
  as a function of the block size $x+L \in [1,2L]$. 
  Left: $h=0.5$, Right: $h=0.05$. The symbols are the numerical results  and
  the continuous lines the theoretical prediction  (\ref{tot}).
  }
\label{fig:compare_entropy}
\end{figure}

As we already mentioned, it is possible to use standard free fermion techniques \cite{ffo} to
calculate the R\'enyi entanglement entropy for an arbitrary block of the rainbow chain, 
as already done in \cite{VRL10,RRS14b,RRS15}.
For a block $A$ starting from the boundary, some results of the numerical computation are reported 
for $n=1,2,3,4$ in Fig. \ref{fig:compare_entropy}.
We only report the results for one total length of the chain equal to $2L=100$ and for two values 
of the inhomogeneity $h=0.5,0.05$, but we have checked that other values of these parameters 
give equivalent results and the CFT predictions are nicely confirmed in their regime of applicability. 
 
Just by looking at Fig. \ref{fig:compare_entropy}, it is clear that the numerical results qualitatively resemble the shape 
of our prediction (\ref{21}) plotted in Fig. \ref{fig:Sblock}. 
However, it is also evident that on top of the smooth conformal contribution, there is an oscillating part 
which becomes more pronounced as the order of the R\'enyi entropy $n$ increases and as $h$ becomes smaller.
Clearly, this effect is not present in the asymptotic result  (\ref{21}), but is a correction to scaling.
These corrections to scaling are well known for homogeneous systems and have been fully characterised 
both for periodic \cite{ccen-10,ce-10} and open systems \cite{lsca-06,FC11}.
In particular in Ref. \cite{cc-10} it has been argued that in CFT they can be understood as 
arising from the insertion of relevant operators locally at the conical space-time singularity which are necessary to describe the 
reduced density matrix and, at half filling in a semi-infinite chain, they are of the form $\sim f_n (-1)^x x^{-1/n}$.
It is then natural to conjecture that, also for the rainbow chain, the leading correction to the scaling is due to this
relevant operator at the conical singularity and so its analytic form is obtained by employing the same conformal 
mapping used for calculating the leading term. 
Following this logic, it is straightforward to arrive to the conjecture for entanglement entropies
\beq
S_n(x) \simeq  S_{n,{\rm CFT}}(x) + S_n^{\rm osc}(x),
\label{tot}
\eeq
where $S_{n,{\rm CFT}}(x)$ is the leading term in  (\ref{21}) while the oscillating part is
\beq
S_n^{\rm osc}(x) = \frac{E_n}{2}+ f_n \cos\left( \pi (x+L) \right)  Y(x)^{ - 1/n}  .
\label{21b}
\eeq
The constants $E_n$ and $f_n$ are non-universal, but, because of the previous argument based on conformal mapping, 
they must be the same as in the homogeneous system. 
Thus, from the exact solution of the homogeneous system in \cite{FC11}, we expect 
\beq
f_n = \frac{2}{1-n} 
\frac{\Gamma \(\frac{1}{2}+\frac{1}{2n}\)}{\Gamma\(\frac{1}{2}-\frac{1}{2n}\)},
\label{21c}
\eeq
while the constants $E_n$ are complicated integrals that we do not report here (see \cite{FC11,jk-04}) 
whose numerical values, for the R\'enyi orders we are interested in, are
\beq
E_ {n=1, \dots, 4} \approx 0.49502, \; 0.40405, \; 0.366365, 0.346061 \, . 
\label{21d}
\eeq

In Fig. \ref{fig:compare_entropy} we compare the prediction  (\ref{tot}) with the numerical results for $2L=100$.
The agreement between the two is excellent and this is particularly amazing since Eq.  (\ref{tot}) does not contain 
any fitting parameter. 
However, at a very close look, it can be noticed a very small offset between data and prediction.
This is a non-oscillating $1/L$ correction which is known in the uniform case (see Eq. (70) in Ref. \cite{FC11}),
but we have been unable to conjecture a generalised form valid for all non-zero $h$. 
We repeated checks like those in Fig. \ref{fig:compare_entropy} for several other chain length $L$ and inhomogeneous 
parameter $h$, always finding excellent agreement.

\subsection{Block $A$ at an arbitrary position} 

We now  consider the more difficult case of a single block of spins but in an arbitrary position in the chain, 
i.e. we consider the bipartition
\beq
A= \[x_1,x_2\],  \quad B = \[-L,x_1\] \cup [x_2,L] ,
\label{24}
\eeq
i.e. $A$ is an interval of length $x_2-x_1>0$. 
Notice that, in the central case $x_1=-x_2$ and in the limit $h\to\infty$, this block has zero entanglement 
given the singlet structure in  Fig. \ref{fig:RB_scheme}.

In CFT formalism, the R\'enyi entropy of this interval  is related to the two-point function of the twist fields $\mathcal{T}_n$,
$\overline{\mathcal{T}}_n$ inside the strip, which can be conformally transformed to a two-point function in the upper half-plane. 
Generically, the calculation of two-point functions of twist fields in the upper half plane is a very challenging problem (in fact, 
by method of images, it can be related to a four point function in the complex plane considered e.g. in Refs. \cite{fps-08,cct-09}). 
However, in the case of the massless free fermionic field-theory, this two-point function in the half plane simplifies and 
it becomes \cite{ch-05,cmv-11,FC11} 
\beq
\fl \< \overline{\mathcal{T}}_n(w_1,\overline{w}_1)
 \mathcal{T}_n(w_2,\overline{w}_2) \>_{\rm uhp}
 \quad = c_n^2 \[ \frac{ \left| w_1 -\overline{w}_2 \right|^2 }
       { \left|w_1 -w_2\right|^2 \left|w_1-\overline{w}_1\right|
         \left|w_2 -\overline{w}_2\right| } \]^{\frac{1}{12}(n-\frac{1}{n})} ,
\eeq
where $c_n$ is the corresponding multiplicative constant of the one point function $\<{\cal T}_n\>$.
This can be conformally mapped back to the strip $(\tilde{x},t) \in [-\tilde{L},\tilde{L}] \times \mathbb{R}$; 
the metric on the strip is, at this point, the euclidean metric $ds^2 = dz d \overline{z}$. 
We only need the result at imaginary time $t=0$, in which case the expression simplifies to
$$
   \< \overline{\mathcal{T}}_n(\tilde{x}_1,0)
  \mathcal{T}_n(\tilde{x}_2,0) \>_{\rm  strip  (flat)} 
 = c_n^2
  \[ \frac{ \frac{1}{4} \( \frac{\pi}{2 \tilde{L}} \)^2
    \cos\(\frac{\pi (\tilde{x}_1+\tilde{x}_2)}{4 \tilde{L}}\)^2 }
     { \sin\(\frac{\pi (\tilde{x}_1-\tilde{x}_2)}{4 \tilde{L}}\)^2
       \cos\(\frac{\pi \tilde{x}_1}{2\tilde{L}} \)
       \cos\(\frac{\pi \tilde{x}_2}{2\tilde{L}} \)} \]^{\frac{1}{12}(n-\frac{1}{n})}.
$$
Finally, we perform the Weyl transformation $dz d \overline{z} \mapsto
e^{2 \sigma} dz d \overline{z} $, which gives the two-point function
we need
$$
  \<\overline{\mathcal{T}}_n(\tilde{x}_1,0)
  \mathcal{T}_n(\tilde{x}_2,0) \>_{\rm  strip (curved)} 
 = c_n^2 \[ 
  \frac{e^{-\sigma(x_1) - \sigma(x_2)} \(\frac{\pi}{4\tilde{L}} \)^2
    \cos\(\frac{\pi (\tilde{x}_1+\tilde{x}_2)}{4 \tilde{L}}\)^2 }
       { \sin\(\frac{\pi (\tilde{x}_1-\tilde{x}_2)}{4 \tilde{L}}\)^2
         \cos\(\frac{\pi \tilde{x}_1}{2\tilde{L}} \)
         \cos\(\frac{\pi \tilde{x}_2}{2\tilde{L}} \)}\]^{\frac{1}{12}(n-\frac{1}{n})}.
$$
The $n^{\rm th}$ R\'enyi entropy is then
\begin{equation}
S_{n,{\rm CFT}}(x_1,x_2) \, =\, \frac{n+1}{12 n} \ln 4Y(x_1,x_2)+E_n,
\label{26} 
\end{equation}
with
\beq
\fl Y(x_1,x_2)  =   
  \frac{ e^{\sigma (x_1) + \sigma(x_2)} 16 \tilde{L}^2 }{ \pi^2 \,
    \cos\(\frac{\pi (\tilde{x}_1+\tilde{x}_2)}{4 \tilde{L}}\)^2 }
  \nonumber
  \sin\(\frac{\pi (\tilde{x}_1-\tilde{x}_2)}{4 \tilde{L}}\)^2
  \cos\(\frac{\pi \tilde{x}_1}{2\tilde{L}} \)
  \cos\(\frac{\pi \tilde{x}_2}{2\tilde{L}} \) ,
\eeq
where, again, $\tilde x_i$ and $\tilde L$ are given by Eqs.  (\ref{5}) and  (\ref{6}) respectively, 
and $E_n$ is given in  (\ref{21d}).
In  (\ref{26}) the factor $4$ and the constant $E_n$ have been fixed in order to match the exact 
result in the homogenous limit $h\to0$ obtained in \cite{FC11}.

\begin{figure}[t]
\centering
\includegraphics[width=0.48\textwidth]{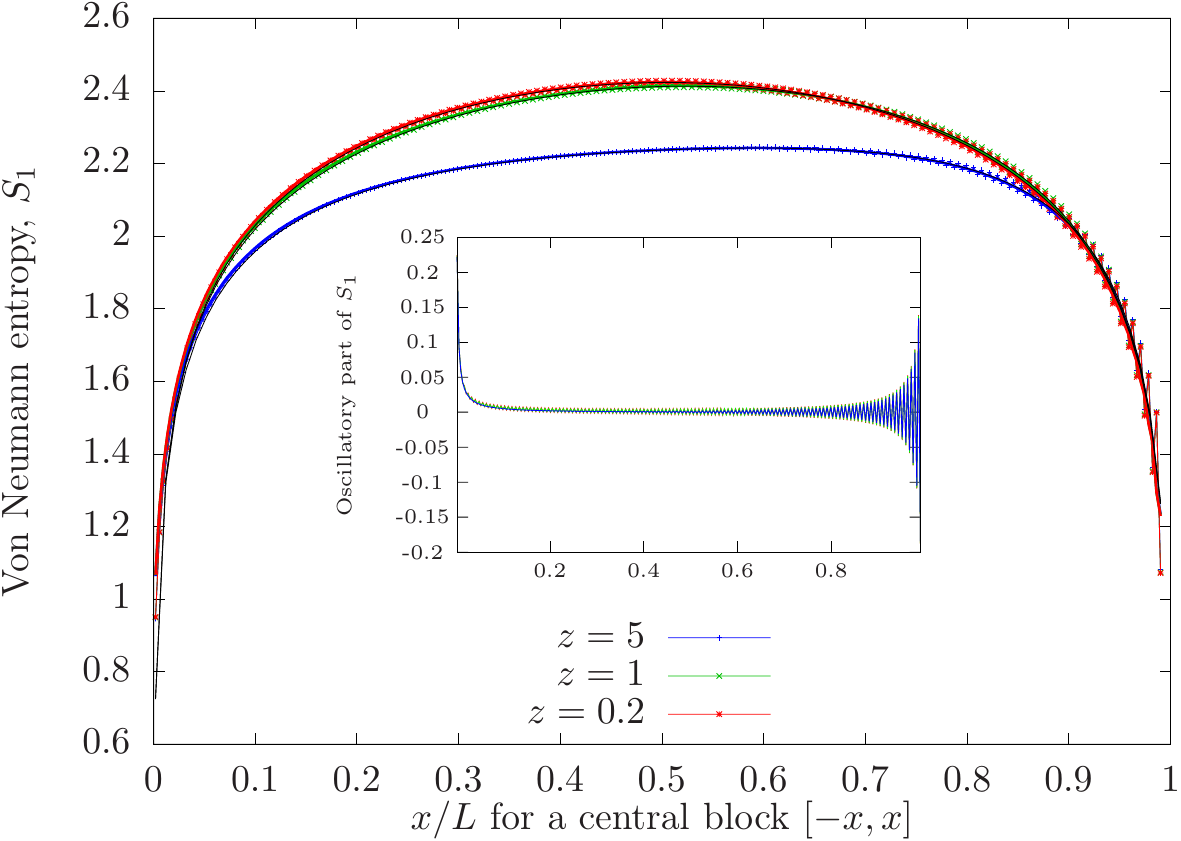}
\hspace{0.2cm}
\includegraphics[width=0.48\textwidth]{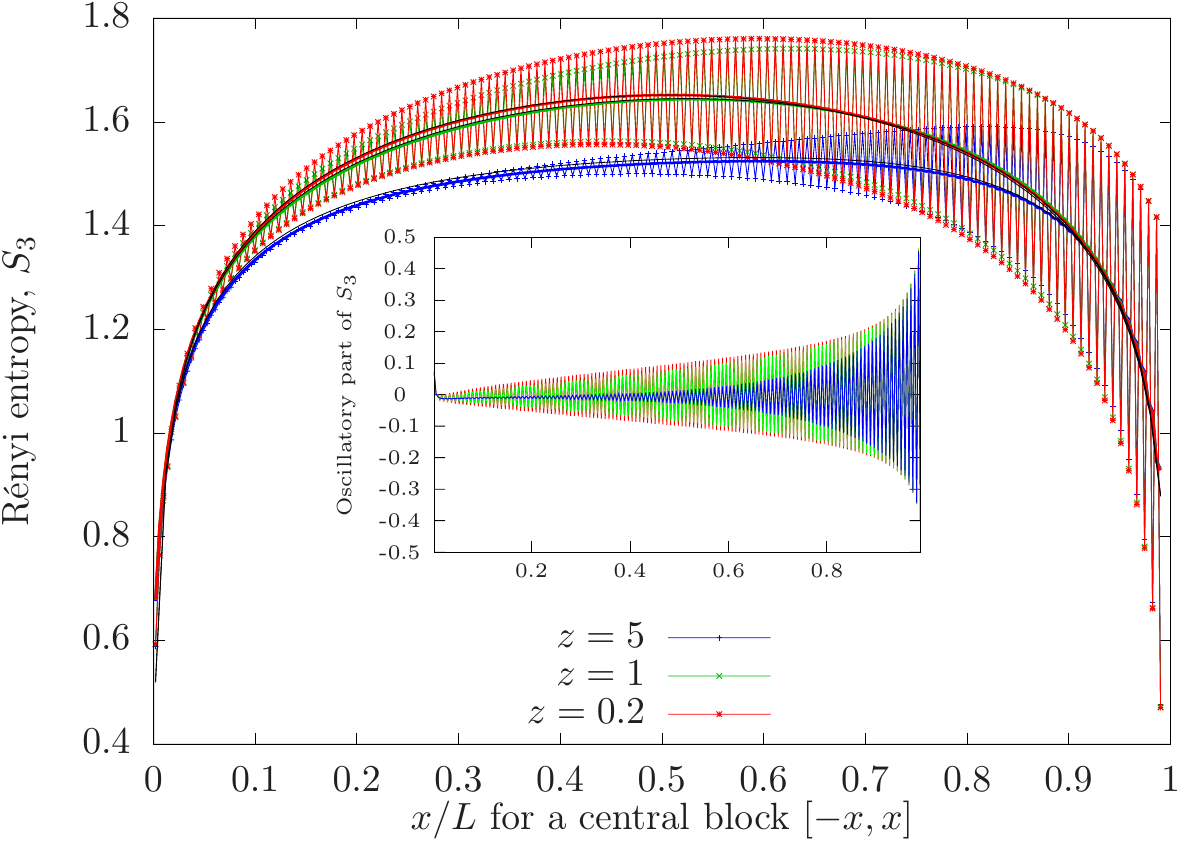}
\caption{Entanglement entropies $S_1(A)$ (left) and $S_3(A)$ (right) for a chain
  with $2L=512$ sites and $z=hL=0.2, 1, 5$. 
  The inset of these figures shows the difference $S_n(x)-S_{n,{\rm CFT}}(x)$, highlighting the irregular behaviour of the 
  oscillatory part. }
\label{fig:compare_central}
\end{figure}

\subsubsection{Numerical tests.}

To be as concise as possible,  we only provide numerical checks of the prediction  (\ref{26}) in the case of a
symmetric block $A = [x_1,x_2] = [-x,x]$ with $x>0$. 
This bipartition is particularly interesting since in the limit $h\to\infty$ the entanglement of this configuration 
vanishes because of the singlet structure (a result that we confirm numerically, but that we do not show).
In Fig. \ref{fig:compare_central} we report the numerically calculated entanglement entropies for this central 
configuration both for $n=1$ and $n=3$ and for three values of $h$ in a regime that is intermediate between 
strong and week inhomogeneity, in order to have non-trivial entropy profiles. 
The CFT prediction  (\ref{26}) is reported together with the numerical data:
the agreement for the average value is really impressive considering that there are no free parameters
in this comparison. 
However, on top of the average value well described by CFT, there are strong oscillations of the form $(-1)^x$ times some function 
of $x$.
These oscillations shares many features with those found in the case of a block starting from the edge, e.g. 
they are more pronounced as $n$ increases, and they decrease with system size.
In order to understand them better, or at least to guess their behaviour, we have studied the difference 
between the numerical $S_n$ and the CFT prediction  (\ref{26}). 
For example, in the inset of Fig. \ref{fig:compare_central}, we report this difference $S_n(x)-S_{n,{\rm CFT}}(x)$ as 
function of the half-block size $x$. 
With system size they are compatible with a scaling $L^{-1/n}$, as expected from general arguments \cite{cc-10}.
However, this scaling is multiplied by a non-trivial function of $x$, that, as shown in the inset of Fig. \ref{fig:compare_central}, grows quite strongly with $x$.
Because of all these complications, we have not been able to obtain (or even guess) an analytic expression for them. 
Their shape resembles the result for the entanglement entropies in the same central configuration found in 
Ref. \cite{clm-15} for a Fermi gas trapped by a harmonic potential. In fact, also in that case it has not been possible 
to provide a result (or a conjecture) for their behaviour. 
We mention also that at a closer look at the data in the inset, is can be noticed that there is a small $n$- and $h$-dependent 
offset in the additive constant. This is expected to be a $1/L$ correction similar to the one observed for the interval starting 
from the edge.

%


\begin{figure}[t]
\centering
\includegraphics[width=9cm]{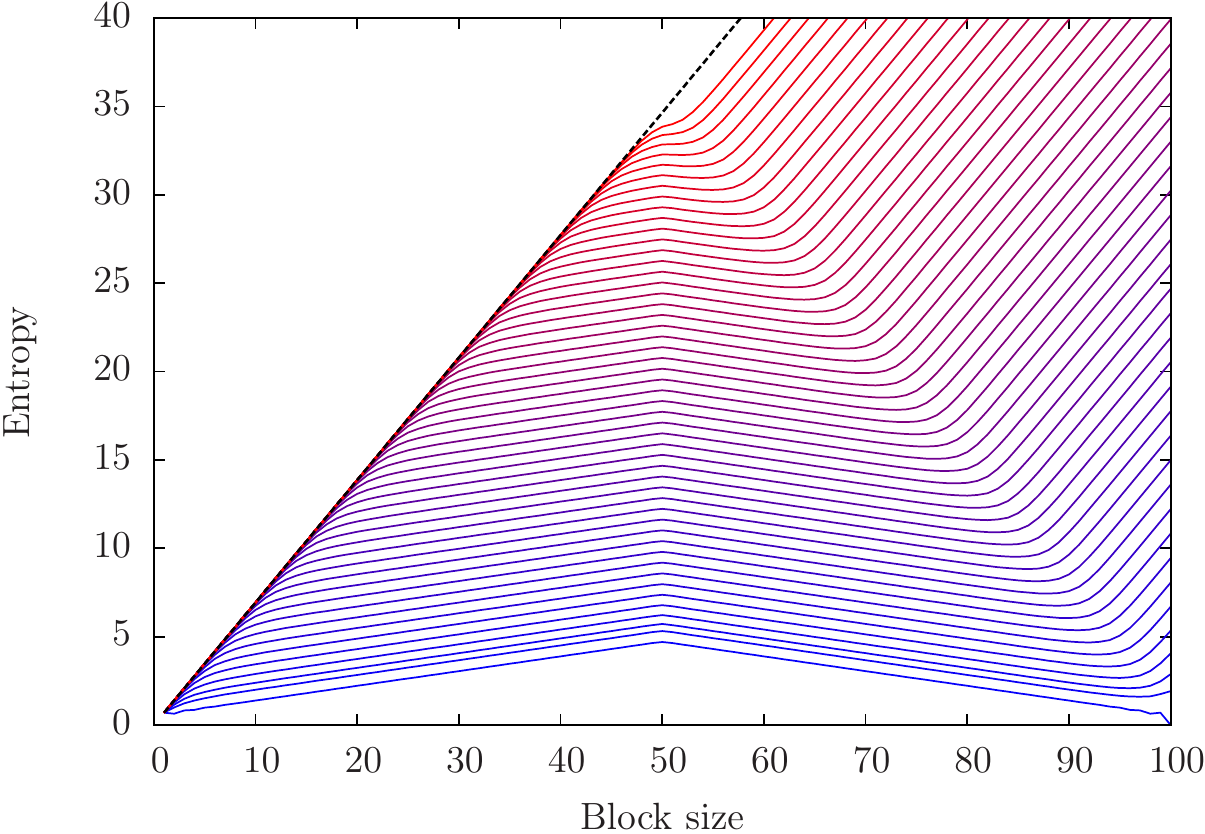}
\caption{Von Neumann entanglement entropy of a rainbow state with $2L=100$ and
  $h=0.5$ (such that $hL=25\gg1$) for different physical temperatures as function of the block size $n$. 
  The colours follow the increasing  temperature from cold (blue) to hot (red), chosen in the following way. 
  The lowest curve is for $T=0$. As we move upwards, the $n$-th curve corresponds to a temperature equal
  to the $n$-th hopping starting from the left as in Eq. (\ref{Tn}). 
  The dashed line on the top corresponds to the maximal possible entropy for each block $S_{\rm max}=n\ln 2$.}
\label{fig:finite_T}
\end{figure}

\section{Rainbow state at finite temperature}
\label{sec:temp}

A final question one naturally poses when studying the rainbow chain  is what happens at finite temperature. 
Given that for $hL\gg1$ the effect of the inhomogeneity is the same as a large temperature, 
it is particularly interesting to address the combined effect of this effective temperature  (\ref{13})
and a real external one. 
Calculating the entanglement entropy at finite temperature in our problem is very difficult. 
Indeed, the topology that one should study is the one of an annulus with two open boundaries in the 
real space coordinate and periodic conditions along the imaginary time.
Calculating either one- or two-point correlation function of twist fields in this geometry is an advanced problem 
that lies beyond the scope of this manuscript.
On the other hand, the most interesting effect we are aiming to describe (i.e. the interplay between real and effective temperature) 
takes place in the limit $hL\gg1$ where the physics considerably simplifies 
and we are able to fully characterise it on the sole basis of  the numerics.

In Fig. \ref{fig:finite_T} we report the effect of the physical temperature on a rainbow chain with $2L=100$ sites for a high value of $h=0.5$ (such that $hL=25\gg1$). 
Fig. \ref{fig:finite_T} shows the von Neumann entropy of blocks starting from the left
boundary, as a function of the block size. 
The set of temperatures, nonetheless, have been chosen with care. 
The lowest curve corresponds to $T=0$, and exhibits the usual tent shape, with a slope
approximately equal to $h/6$ (recall Eq.  (\ref{11})). 
Above this curve, the $n$-th one corresponds to a temperature equal to the $n$-th
hopping, counting from the left end, i.e. 
\beq
T_n = e^{-h (L-1/2-n)}, \qquad n=1,\dots, L-1.
\label{Tn}
\eeq
The dotted curve corresponds to the maximal possible entropy for each block size, that is $n\,\ln 2 $
(in this section we only consider the von Neumann entropy and so we denote the site with $n$
without risking to make confusion with the R\'enyi index).

\begin{figure}[t]
\includegraphics[width=8cm]{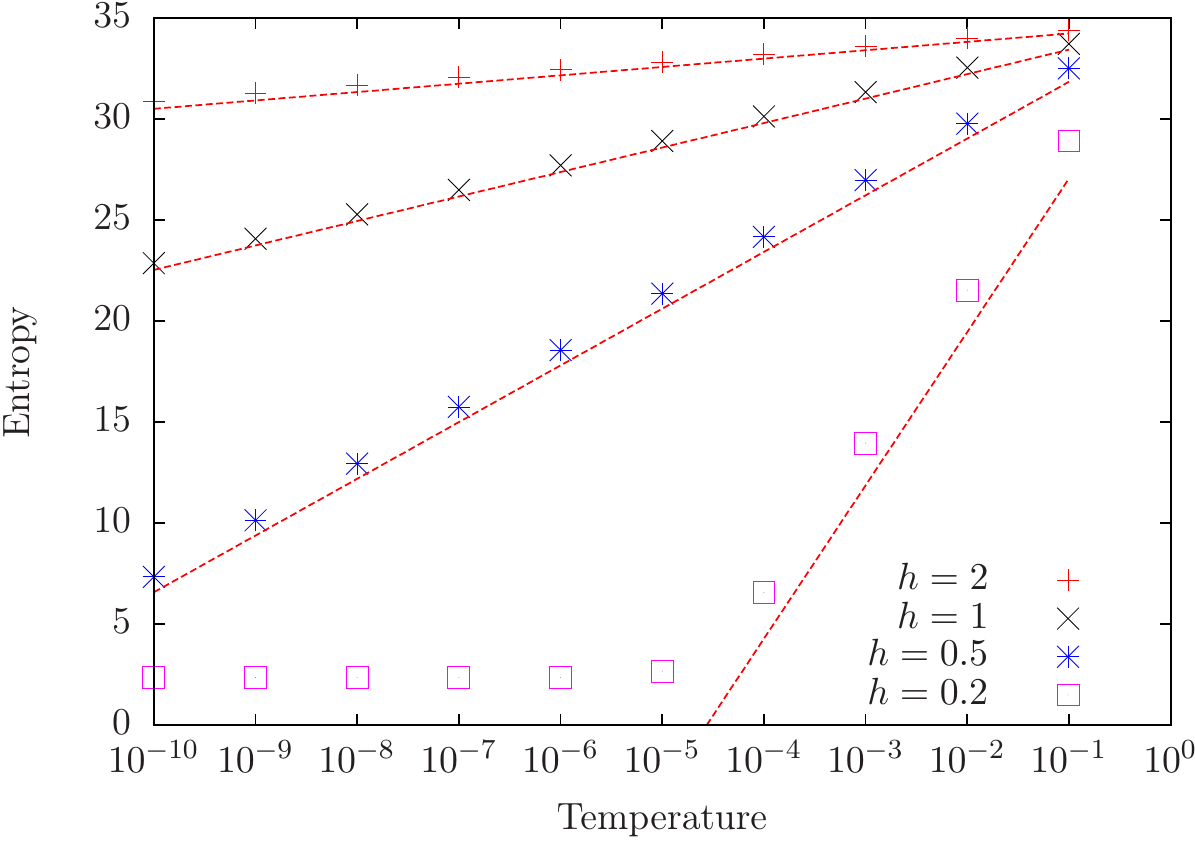}
\caption{Entanglement entropy of the half chain as a function of the temperature
  for rainbow chains with $2L=100$ and different values of $h$. 
  The dashed red line is the theoretical approximation  (\ref{eq:entropy_finiteT_halfsize}). 
  Notice that the approximation  is correct as long as the temperature lies within the range of hoppings. 
  For $h=0.2$ it fails for $T\approx 10^{-5}$, because these temperatures are below the minimum hopping.}
\label{fig:entropy_temp_half}
\end{figure}

The results in Fig. \ref{fig:finite_T} show that for fixed finite temperature there are three 
linear behaviours (except at the turning points where some crossover takes place quite rapidly) 
with slopes that can be explained in the following way. 
Let us assume that we set the physical temperature
equal to the hopping at a certain point $x_0>0$ (and its reflection,
$-x_0$): $T=J(x_0)$. The blocks lying on the left edge, i.e. $(-L, x)$
with $x < -x_0$, involve hopping amplitudes much lower than the
temperature, that is $J(x)\ll T$, and hence the entropy will be
maximal and given approximately by $(L- |x|) \ln 2$. When the right
point of the block $(-L, x)$ moves past $-x_0$, now the local hopping
is much larger than the temperature, $J(x)\gg T$, and the behaviour of
the entropy is as in the zero-temperature case. Finally, when $x >
x_0$, one gets $J(x)\ll T$, and the entropy growth is again the
maximal one. Summarising, we have 
\begin{equation}
S(x) \sim
\left\{ 
\begin{array}{ll} 
(L- |x|) \ln 2,  & x \in (- L, - x_0), \\
(L- x_0) \ln 2 + (x_0 - |x|) h/6 , & x \in ( - x_0, x_0), \\ 
(L- 2 x_0 +x ) \ln 2,  & x \in (x_0, L). \\
\end{array}
\right. 
\label{eq:approx_entropy}
\end{equation}
In other terms: if the right extreme of the block is in the interval
$[-x_0,x_0]$, the behaviour will be similar to the zero temperature
case, and if it is outside that region, it will correspond to infinite temperature. 
Thus, the various curves can be just characterised by the value they have at $x=0$, i.e. by the half-chain entropy.
According to the previous argument, we have 
for a rainbow system with a given $h$ (such that $hL\gg1$) and physical temperature $T$, the half-chain entropy
(recall $T = J(x_0) = e^{ - h x_0}$) 
\begin{equation}
S (x=0)  \simeq \( \frac{\ln 2}{h} - \frac{1}{6} \) \ln T   + L \ln 2 .
\label{eq:entropy_finiteT_halfsize}
\end{equation}
The correcteness of this prediction is checked in Fig. \ref{fig:entropy_temp_half}. 
The range of validity of the approximation is for $T$ within the range of values
for the hopping, i.e. if it exists $x_0$ such that $T=J(x_0)$. Notice that, for $h=0.2$, this range is more reduced.

\section{Conclusions}
\label{ref:conclusions}

In this paper we applied methods of 2d quantum field theory in curved
space-time to the study of the {\em rainbow chain},
which has a ground state that for $hL\gg1$ possesses
a volume law, analogous to a thermal state with large temperature $T_R=h/(2\pi)$.

We first showed  that the rainbow system is describable by a massless Dirac fermion on
a Riemannian manifold with constant negative curvature everywhere except at the centre, equivalent to a Poincar\'e metric with 
a strip removed.
This identification allowed us to apply
the recent results of Ref. \cite{DSVC16} to the rainbow state. 
In this way, we provided accurate predictions for the smooth part of the entanglement entropies of blocks of different types
which perfectly describe the numerical data. 
We also conjectured a precise form for the oscillatory part of the entropies in the case of edge blocks. 

The dynamics of the rainbow state at finite temperature is also very peculiar. 
For $T$ in the range of energies spanned by the values of the hopping amplitude,  we showed that the system splits into three
regions: the central one behaves as if it were at $T=0$, while the two extremes  as if they were at $T=\infty$.

There are still many more results that can be derived by using the curved CFT approach and that can help understand the 
structure of the rainbow model. 
For examples, it should be easy to generalise the results presented here for the ground state to excited states, both 
low lying ones and those in the middle of the spectrum. 
For the former, one can use the universal CFT approach \cite{abs-11,txas-13}, while for the latter one 
can exploit the results in \cite{afc-09}.
Another possible generalisation is to consider other entanglement measures that have a CFT representation 
such as bipartite fidelity \cite{ds-11}, entanglement negativity \cite{CCT}, and relative entropies \cite{lash-14}. 

Moving away from the rainbow, the dynamics of conformal fields on curved backgrounds provides a rich
source of physical phenomena to explore, whose interest is growing due
to the recent proposals of experimental realisations in, e.g.,
ultracold atoms on optical lattices \cite{BCLL11,RTLC16}. The subtle
interplay between entanglement and geometry helps explain the thermal
effects in quantum field theory on curved backgrounds, e.g. the Unruh
effect or the local temperature effects in inhomogeneous fermionic
systems \cite{RR16}.


\section*{Acknowledgements}

We would like to thank A. Abanov, L. Alvarez-Gaum\'e, V. Korepin, E. L\'opez, J. Mas,
G. Mussardo, E. Tonni, J. Pachos, A. Ramallo and T. Takayanagi for
useful comments. This work was funded by grants FIS-2012-33642,
FIS-2012-38866-C05-1, and FIS2015-69167-C2-1-P from the Spanish
government, QUITEMAD+ S2013/ICE-2801 from the Madrid regional
government and SEV-2012-0249 of the Centro de Excelencia Severo Ochoa
Programme. J.D. and G.S. thank Nordita, Stockholm, for hospitality
during the Workshop ``From quantum field theories to numerical
methods'', where part this work was done. J.D. acknowledges financial
support from CNRS ``D\'efi Inphyniti'' and from the Conseil R\'egional
and Universit\'e de Lorraine.
G. S. thanks SISSA for hospitality during the starting stage of this project.
P.C. acknowledges funding from the ERC under Starting Grant 279391 EDEQS.


\end{document}